\documentclass[a4paper,11pt]{article}
\pdfoutput=1 

\usepackage{jcappub}

\usepackage[T1]{fontenc}

\title{Higher derivative theories with constraints : exorcising Ostrogradski's ghost}

\author[a,1]{Tai-jun Chen,\note{Corresponding author.}}
\author[b]{Matteo Fasiello}
\author[a,c]{Eugene A. Lim}
\author[b]{Andrew J. Tolley}

\affiliation[a]{DAMTP, University of Cambridge, Wilberforce Road, CB3 0WA, Cambridge, U.K.}
\affiliation[b]{Case Western Reserve University, 2076 Adelbert Road, Cleveland Ohio, 44106, U.S.A.}
\affiliation[c]{Theoretical Particle Physics and Cosmology Group, Physics Department, King's College London, Strand, London, U.K.}

\emailAdd{T.Chen@damtp.cam.ac.uk}
\emailAdd{matte@case.edu}
\emailAdd{eugene.a.lim@gmail.com}
\emailAdd{andrew.j.tolley@case.edu}

\abstract{We prove that the linear instability in a non-degenerate higher derivative theory, the Ostrogradski instability, can only be removed by the addition of constraints if the original theory's phase space is reduced.}

\begin{document}
\maketitle
\flushbottom

\section{Introduction}

When Newton wrote down his second Law of Motion
\begin{equation}
\ddot{q} = \frac{F(q)}{m},
\end{equation}
i.e. motion is described by an equation second order in the time derivative of the fundamental dynamical variable position $q$, he chose wisely. As it is now well known, almost two hundred years later, Ostrogradski \cite{Ostro} proved a theorem which showed that in any non-degenerate theory whose fundamental dynamical variable is higher than 2nd order in time derivative there exist a linear instability.

Consider instead, if Newton had chosen the fourth order theory
\begin{equation}
 q^{(4)}= \frac{d\alpha}{dq} \label{eqn:4thorderNewton}
\end{equation}
with, $\alpha$ being some function of $q$, i.e. a potential. This equation of motion can be obtained from a higher derivative action of the following form:
\begin{equation}
S = \int dt~\left(\frac{1}{2}\ddot{q}^2 - \alpha(q)\right). \label{eqn:S1}
\end{equation}
Since eq.(\ref{eqn:4thorderNewton}) is 4th order, the phase space is 4 dimensional. Without going too much into details at the moment --- we will get there soon enough --- we can describe the phase space by a pair of canonical variables and their momenta $(P_1,Q_1)$ and $(P_2,Q_2)$, with the Hamiltonian
\begin{equation}
H = P_1 Q_2 + \frac{P_2^2}{2} + \alpha(Q_1). 
\end{equation}
One can always choose $\alpha(Q_1)$ to be some function which is bounded from below, say $\alpha = Q_1^2$. More problematic, however, is the first term which signals the famous Ostrogradski linear instability. The ``linear'' in ``linear instability'' refers to the linearity of the $P_1$ in this term --- since $P_1$ is free to roam the phase space, there is no barrier that prevents some degrees of freedom of the theory from probing arbitrarily negative energies. In other words, the Hamiltonian is not bounded from below.\footnote{Technically, it is single side boundedness that is important, a Hamiltonian that is bounded from above is equally as good --- one can simply flip the sign of the Hamiltonian.}

This instability  on its own is not a bad thing. It becomes bad when interactions with other degrees of freedom whose Hamiltonians are bounded from below are introduced. The presence of these negative energy states means that there exist a vast phase space where the Hamiltonian is negative, hence the modes will begin to populate them by entropic argument alone while, by conservation of energy, creating an equally large number of positive energy modes in the interacting d.o.f. \cite{Kallosh:2007ad,Eliezer:1989cr}. This is the onset of the instability. Note that, while this is a \emph{classical} instability, in quantum theory, negative energy modes are particularly sick --- attempts to canonically quantize them will either lead to negative norm states (and hence undefined) or negative energy states (and hence runaway particle production). Since negative norm states are often called ``ghosts'' in quantum theory, higher derivative theories are often called ``ghost-like''.\footnote{However, in some parameter space the \textquotedblleft ghosts\textquotedblright become  \textquotedblleft tachyons\textquotedblright , one can always check it from the wavefunction of the theory. If the wavefunction is oscillatory (exponentially growing/decaying), it is a ghost (tachyon).}

Recently, there has been a resurgence of interest in higher derivative theories, particularly within attempts to modify gravity \cite{Myers:1987yn,Charmousis:2012dw,Clifton:2011jh,Moldenhauer:2010zz,Buchdahl:1983zz,Stelle:1977ry,Calcagni:2005im,Hindawi:1995an,Chiba:2005nz,Nunez:2004ts,Barth:1983hb,Mannheim:2011ds}. It is well known that higher derivative theories of the $f(R)$ form are secretly healthy as they are \emph{degenerate} --- a technically important distinction which means that the highest derivative term cannot be written as a function of canonical variables. In fact, $f(R)$ can be recast as an (interacting) theory of a scalar and two graviton modes \cite{Strominger:1984dn,Magnano:1993bd,Biswas:2005qr, Sotiriou:2008rp, DeFelice:2010aj,Woodard}. On the other hand, theories employing curvature invariants such as  $R_{\mu\nu}R^{\mu\nu}$, $R_{\mu\nu\sigma\gamma}R^{\mu\nu\sigma\gamma}$ or the Weyl invariant $C_{\mu\nu\sigma\gamma}C^{\mu\nu\sigma\gamma}$ \cite{Mannheim:2011ds,Maldacena:2011mk,Lu:2012xu}, are \emph{non-degenerate} higher derivative theories. Hence they suffer from the sickness of Ostrogradski instability. 

Furthermore, there is also great interest in the so-called ``higher derivative'' scalar field theories, such as the Galileon or Lovelock gravity \cite{Lovelock:1971yv,Nicolis:2008in,Deffayet:2009wt, Deffayet:2010qz,Deffayet:2009mn} which, when coupled non-trivially to the metric, result in interesting scalar field dynamics which cannot be reproduced by simple $f(R)$-type modifications. These theories, while naively looking like ``higher derivative'' theories (in the sense that in the Lagrangian there are terms of 2nd order and higher in time derivatives), are secretly completely healthy \emph{non}-higher-derivative theories; their E.O.M. are 2nd order in time derivatives and so are dimension $2$ in phase space. These properties have been achieved by the addition of structure in the Lagrangian --- usually by the clever cancellation of higher derivative terms in the E.O.M.. We do not consider this class of theories as higher derivative theories and they do not suffer from the instability.

On the other hand, in true non-degenerate higher derivative theories, the instability is ubiquitous --- as we will review below (also see \cite{Woodard,Simon:1990ic}). Since Ostrogradski's theorem is so simple to prove and requires very little initial assumptions, it is incredibly powerful \cite{Kallosh:2007ad, Jaen:1986iz, deUrries:1998bi, Aref'eva:2006xy, Hinterbichler:2011tt,deRham:2012az} (see also \footnote{However, it is claimed that some theories with infinite order time derivatives (i.e. the nonlocal theory) are free of Ostrogradski's instability. See \cite{Woodard, Eliezer:1989cr, Barnaby:2006hi}}).

One possible way out is to impose boundary conditions or initial conditions such that the unstable modes are zero. For example, in \cite{Maldacena:2011mk}, the higher derivative terms in the Weyl term $C_{\mu\nu\sigma\gamma}C^{\mu\nu\sigma\gamma}$ are ``removed'' by imposition of such boundary conditions. However, this is not a satisfactory solution: as we explained above in the simple example with Newton's law of motion, the presence of \emph{any} interaction will immediately source these modes and the instability will inevitably build up.

Another interesting way out is suggested by Bender and Mannheim in \cite{Bender,Mannheim} where they show that ${\cal PT}$-symmetric Hamiltonians are ghost-free up to fourth order, although this entails giving up the Hermiticity of the Hamiltonian. In addition, since the classical theory in this case still possesses negative energies, the Correspondence Principle is abandoned and the interpretation of this formalism is unclear.

Finally, one might try to eliminate the instability by imposing constraints (for example, those suggested by \cite{Eliezer:1989cr, Durrer:2008in,Lim:2010yk}), i.e. one selectively restricts the trajectories of the d.o.f. such that the Hamiltonian becomes bounded from below. 

The implementation of constraints into the theory requires the introduction of auxiliary variables and hence the enlargement of the total phase space (the dimensionality of the \emph{reduced} phase space is still the same or smaller since trajectories are constrained). As a consequence, one may hope to change the orbits of the trajectories of the theory to a degree which is sufficient to cure it of the instability. 

Using our fourth order theory example above, one can imagine a modification
\begin{equation}
S = \int dt~\left(\frac{1}{2}\ddot{q}^2 - \alpha(q) + \lambda f(q,\dot{q},\ddot{q})\right), \label{eqn:S2}
\end{equation}
where $\lambda$ is an auxiliary field which enforces the constraint $f(q,\dot{q},\ddot{q})=0$. We emphasize that the action eq. (\ref{eqn:S2}) is a \emph{different} physical theory from the action eq. (\ref{eqn:S1}) as long as the constraint cannot be gauged away. Can we cleverly choose $f$ such that this theory, despite being a higher derivative theory, is free of the linear instability? 

In this paper we will prove that in order to remove the instability by the imposition of constraints, the constraints must \emph{reduce} the effective dimensionality of the phase space of the original theory. For example, an unstable theory with a phase space dimensions of six can be rendered stable by reducing the phase space to dimension four or less employing auxiliary or Lagrange multiplier fields.

The layout of the paper is as follows. In section \ref{sect:example} we review Ostrogradski's Theorem. In section \ref{sect:donotsave} we show that, for the simple case of a $2$nd order (in the action) theory, the addition of Lagrange multipliers which do not reduce the original phase space leaves the theory unstable. In section \ref{sect:generalproof} we prove in general the previous statement. In section \ref{sect:remove}, we show how an unstable theory can be rendered stable by reducing the dimensionality of the original phase space. In section \ref{sect:summary} we will summarize.

\section{Ostrogradski's theorem: an example} \label{sect:example}
Ostrogradski's theorem \cite{Ostro, Woodard} can be stated as follows:

\emph{If the higher order time derivative Lagrangian is non-degenerate, there is at least one linear instability in the Hamiltonian of this system.}

\emph{Non-degeneracy} means that the highest time derivative term can be expressed in terms of canonical variables. In the usual first order theory with a single degree of freedom $q$, this means that we can express the dynamical variable $\dot{q}$ as a function of canonical variables $Q$ and $P$. For example, in a theory with $L = (1/2)\dot{q}^2 - (1/2)q^2$, the canonical momentum is
\begin{equation}
P = \frac{\delta L}{\delta \dot{q}}  = \dot{q}
\end{equation}
which we can then trivially invert $\dot{q} = F(P,Q) = P$. In a higher derivative theory, this translates into expressing $q^{(N)}$ as a function of the canonical variables $Q_i$ and $P_i$. Degenerate theories, on the other hand, are non-invertible and are either stable on their own or  may be made stable with the introduction of constraints \cite{Woodard} --- we will not discuss such theories in this paper. 

A famous example of a higher derivative non-degenerate theory is the Pais-Uhlenbeck (PU) oscillator \cite{Pais}. Here we follow the discussion in ref. \cite{Bender,Mannheim}. The PU action is given by 
\begin{equation}
S_{PU}=\int dt L_{PU}=\frac{\gamma}{2}\int dt[\ddot{q}^2-(w_1^2+w_2^2)\dot{q}^2+w_1^2w_2^2q^2]
\end{equation}
where $\gamma$, $w_1$, and $w_2$ are positive constants and without loss of generality we take $w_1\geq w_2$. The equation of motion of the Pais-Uhlenbeck oscillator contains terms up to the fourth time derivative:
\begin{equation}
\frac{d^4 q}{dt^4}+(w_1^2+w_2^2)\frac{d^2q}{dt^2}+w_1^2w_2^2q=0,
\end{equation}
hence requiring four initial value data ($q_0$, $\dot{q}_0$, $\ddot{q}_0$, $q_0^{(3)}$), allowing us to solve for $q(t)$, to obtain:
\begin{eqnarray}
q(t)&=&-\frac{w_2^2q_0+\ddot{q}_0}{w_1^2-w_2^2} \cos(w_1 t)-\frac{w_2^2\dot{q}_0+q_0^{(3)}}{w_1(w_1^2-w_2^2)} \sin(w_1 t)\nonumber \\
&& +\frac{w_1^2q_0+\ddot{q}_0}{w_1^2-w_2^2} \cos(w_2 t)+\frac{w_1^2\dot{q}_0+q_0^{(3)}}{w_2(w_1^2-w_2^2)} \sin(w_2 t). \nonumber \\
&& 
\end{eqnarray}

Since the solution depends on four initial value data, the phase space must be four dimensional, and Ostrogradski's choice for the canonical coordinates is
\begin{eqnarray}
Q_1\equiv q  &\longleftrightarrow& P_1\equiv \frac{\delta L_{PU}}{\delta \dot{q}}=-\gamma(w_1^2+w_2^2)\dot{q}-\gamma q^{(3)}  \nonumber \\
Q_2\equiv\dot{q}   &\longleftrightarrow&  P_2\equiv \frac{\delta L_{PU}}{\delta \ddot{q}}=\frac{\partial L_{PU}}{\partial\ddot{q}}=\gamma \ddot{q}. \nonumber
\end{eqnarray}
Non-degeneracy implies that $\ddot{q}$ can be inverted and written as a function of the canonical variables $Q_i$ and $P_i$ --- here this is clearly the case. On the other hand, a degenerate model is always guaranteed to have constraints. For example, if the model is degenerate, say if  $P_2 = \delta L_{PU}/\delta \ddot{q}$ is arbitrary function $f(q, \dot{q})$ but not of $\ddot{q}$, then from the definition of $P_2$ there will be a primary constraint $P_2-f(Q_1,Q_2)=0$, which will reduce the number of physical degrees of freedom and the final phase space will be smaller. 

The Hamiltonian of Pais-Uhlenbeck is as usual obtained by Legendre transforming
\begin{eqnarray}
H_{PU}&=&P_1 \dot{q}+P_2 \ddot{q}-L_{PU}\nonumber \\
&=&P_1Q_2+\frac{P_2^2}{2\gamma}+\frac{\gamma}{2}(w_1^2+w_2^2)Q_2^2-\frac{\gamma}{2}w_1^2w_2^2Q_1^2\label{6}.\ \ \ \ \ \ \
\end{eqnarray}

The Hamiltonian generates the time evolution of any function of canonical variables $F(Q_i, P_i)$ via the Poisson Bracket $\dot{F}(Q_i, P_i)=[F(Q_i, P_i), H_{PU}]_P$. One can check that the evolution equations in this Hamiltonian formalism reproduce the Euler-Lagrange equation in the Lagrangian formalism, so it is the right Hamiltonian of the system. The Hamiltonian is conserved if the Pais-Uhlenbeck Lagrangian is not explicitly dependent on $t$, thus we can as usual view the Hamiltonian as ``energy". \\

As in eq. (\ref{6}), the Hamiltonian is linearly dependent on $P_1$ and it means the system is unstable. The $P_1Q_2$ term can be arbitrarily negative when $P_1\rightarrow -\infty$. The Hamiltonian is thus unbounded from below, which means that there is no well-defined vacuum state in the theory. Ostrogradski's result is that \emph{all} the Hamiltonians of non-degenerate higher time derivative theory suffer from linear instabilities. 

\section{Constraints do not cure Ostrogradski's instability if the dimensionality is not reduced} \label{sect:donotsave}
In this section, we will show that the Ostrogradski's instability in general cannot be cured by adding constraints into the theory if the dimensionality of phase space is not reduced by the constraints, i.e. one can only possibly selectively constrain the unstable degrees of freedom and remove them from the physical phase space if the dimension of the phase space is less than that of the original unstable higher derivative theory. We will introduce the constraints by auxiliary variables $\lambda_i$ in such a way that there is no time derivative on $\lambda_i$ in the Lagrangian and primary constraints are thus introduced through their canonical momenta, $P_{\lambda_i}=0$. We follow Dirac's method to analyze the higher order theory with constraints \cite{Dirac,Henneaux,Gitman,Prokhorov,Govaerts}. 

First we will show that the most general non-degenerate second time derivative Lagrangian with one extra auxiliary field (and hence a pair of second class constraints) does not cure its instability without the dimensionality of phase space being reduced. We then apply this result to the Pais-Uhlenbeck model. We  generalize our result to any $N$-th order non-degenerate higher derivative theory in section \ref{sect:generalproof}.

\subsection{First vs second class constraints}
We pause here to introduce the notion of \emph{first class} and \emph{second class} constraints. Second class constraints are ``physical'' in the sense that the solutions to the equations of motion are different with or without the constraint --- e.g. a train restricted to move on a fixed railroad which enforces the second class constraint. On the other hand, first class constraints are those associated with some gauge freedom in the theory, i.e. the solutions of the equations of motion contain some arbitrary functions of time and hence describe physically equivalent systems.

As we know, one can ``gauge fix'' such theories --- these so-called ``gauge fixing'' functions appear as new (primary) constraints in the theory, and once introduced the original first class constraint and the new gauge fixing constraint both become second class constraints.  Hence, when considering the instability, it is clear that once a general proof for second class constraints is shown, it is complete --- physics does not depend on gauge choices after all.

\subsection{General second order non-degenerate theory with second class constraint}
The most general second order time derivative Lagrangian with one auxiliary field $\lambda$ is given by the Lagrangian
\begin{equation}
L=f(q, \dot{q}, \ddot{q}, \lambda).
\end{equation}
The equations of motion of this Lagrangian are
\begin{align}
\frac{\partial f}{\partial\lambda}&=0  \\
\frac{\partial f}{\partial q}-\frac{d}{dt}\left(\frac{\partial f}{\partial\dot{q}}\right)+\frac{d^2}{dt^2}\left(\frac{\partial f}{\partial\ddot{q}}\right)&=0\label{9}, 
\end{align}
where the assumption of non-degeneracy means that the eq. (\ref{9}) is a fourth order differential equation and we can solve $q(t)$ and $\lambda(t)$ with six initial value data ($q_0$, $\dot{q}_0$, $\ddot{q}_0$, $q^{(3)}_0$, $\lambda_0$, $\dot{\lambda}_0$). The phase space is thus six-dimensional and, following Ostrogradski's spirit, the choice of canonical variables is
\begin{eqnarray}
Q_1\equiv q \ \ \ \  &\longleftrightarrow& \ \ \ P_1\equiv \frac{\delta L}{\delta \dot{q}}=-\frac{d}{dt}\frac{\partial f}{\partial \ddot{q}}+\frac{\partial f}{\partial \dot{q}}\ \ \ \ \ \ \ \  \\
Q_2\equiv\dot{q} \ \ \ \  &\longleftrightarrow& \ \ \ P_2\equiv \frac{\delta L}{\delta \ddot{q}}=\frac{\partial f}{\partial\ddot{q}}\label{11}\ \ \ \ \ \ \\
Q_3\equiv \lambda \ \ \ \  &\longleftrightarrow& \ \ \ P_3\equiv \frac{\delta L}{\delta \dot{\lambda}}=0,\ \ \ \ \ \ \ \  
\end{eqnarray}
where $\Phi_1: P_3=0$ is the primary constraint, here we introduce the notation `:' to denote  ``functional form given by''. The assumption of non-degeneracy allows us to use eq.(\ref{11}) to invert $\ddot{q}= h(Q_1, Q_2, Q_3, P_2)$. The total Hamiltonian $H_T$ of this system is 
\begin{eqnarray}
H_T&=&P_1Q_2+P_2h(Q_1, Q_2, Q_3, P_2)  -f(Q_1,Q_2,Q_3,h) \nonumber \\
&& +u_1\Phi_1
\end{eqnarray}
where $u_1$ is a function of canonical variables which can be found later,\footnote{That is, by imposing the consistency relations.} but since we are only interested in the stability of the physical degrees of freedom on the reduced phase space, we will not write $u_1$ explicitly. 

Since $P_3=0$, consistency implies that its equation of motion $\dot{P_3}=[P_3,H_T]$ must also vanish (on constraint) --- this leads to a series of \emph{consistency relations} which allow us to find further constraints called secondary constraints. In this case, there exist one further secondary constraint as expected, which is
\begin{eqnarray}
\Phi_2: [\Phi_1, H_T]_P&=&-P_2\frac{\partial h}{\partial Q_3}+\frac{\partial h}{\partial Q_3}\left[\frac{\partial f}{\partial \ddot{q}}\right]_{\ddot{q}=h}+\left.\frac{\partial f}{\partial \lambda}\right|_{\lambda=Q_3}\nonumber \\
&=& \left.\frac{\partial f}{\partial \lambda}\right|_{\lambda=Q_3}(Q_1, Q_2, Q_3, h)\approx0. \label{eqn:cons1}
\end{eqnarray}
Here we introduce the \textit{weak equality symbol} ``$\approx$" for the constraint equations. The constraint equation is written as $\Phi_2\approx 0$, which means $\Phi_2$ is numerically restricted to be zero but does not identically vanish throughout phase space. I.e. $\Phi_2$ only vanishes on the hypersurface where all the constraints are satisfied.

If $\Phi_2$ is dependent on $Q_3$, then both $\Phi_1$ and $\Phi_2$ are second class constraints and there are no further constraints from the consistency relations; further consistency relations only tell us the form of the arbitrary function $u_1$. Using the two second class constraints we can rewrite $(Q_3, P_3)$ as functions of other canonical variables ($Q_3\approx \mathcal{G}(Q_1, Q_2, P_2)$, $P_3=0$). The reduced Hamiltonian $H_R$ of the physical degree of freedom becomes
\begin{eqnarray}
H_R&=&P_1 Q_2+P_2 h(Q_1, Q_2, \mathcal{G}(Q_1, Q_2, P_2), P_2)\nonumber\\
&& -f(Q_1, Q_2, h, \mathcal{G}).
\end{eqnarray}
The reduced Hamiltonian is always linearly dependent on $P_1$ for any conceivable Lagrangian $L=f(q, \dot{q}, \ddot{q},\lambda)$, which is the signal of instability. 

On the other hand, if $\Phi_2$ is not dependent on $Q_3$, $\Phi_1$ and $\Phi_2$ commute with each other. In this case, we can find further constraints from the consistency relations (and hence a reduction in the effective dimensionality of the phase space)  and we should check whether the constraints are first or second class after we find all the constraints. We will give examples in the following sections.

Even in this simple example, one can quickly see the only possibility to cure the instability comes from the further constraints generated by consistency relation ($\Phi_3$ and $\Phi_4$). The instability's root cause is the pesky linear term $P_1 Q_2$, and to fix this instability one must find a constraint where $Q_2$ must be some function of $P_1$ --- but it is clear when generating the constraint $\Phi_2$ with the consistency relation eq. (\ref{eqn:cons1}), $P_1$ never enters the equation.

\subsection{Pais-Uhlenbeck model with constraint}
We will now apply the above result to the Pais-Uhlenbeck model as an example. We consider the constraint $\ddot{q}=\dot{q}$ to present a flavor of how the instability cannot be evaded if the dimensionality is not reduced. In this case, the dimensionality of the phase space remains the same (i.e. 4) with or without the constraint term. 

\subsubsection{Constraint: $\ddot{q}^2-\dot{q}^2=0$} \label{sect:PUqddot}
The Lagrangian of Pais-Uhlenbeck model with constraint $\ddot{q}^2-\dot{q}^2=0$ is given by
\begin{equation}
L_{PUC}=\frac{\gamma}{2}[\ddot{q}^2-(w_1^2+w_2^2)\dot{q}^2+w_1^2w_2^2q^2]+\frac{\lambda}{2}(\ddot{q}^2-\dot{q}^2).
\end{equation}
This model is an example where $\Phi_2$ is dependent on $Q_3$. 

The equations of motion from varying with respect to $\lambda$ and $q$ become differential equations of both variables
\begin{eqnarray}
\ddot{q}^2-\dot{q}^2&=&0  \\
\gamma\frac{d^4 q}{dt^4}+\gamma(w_1^2+w_2^2)\frac{d^2q}{dt^2}  \nonumber \\
+\gamma w_1^2w_2^2+\frac{d^2}{dt^2}(\lambda \ddot{q})  + \frac{d}{dt}(\lambda \dot{q})&=&0,
\end{eqnarray}
and the functions $q(t)$ and $\lambda(t)$ can be solved with four initial value data $q_0$, $\dot{q}_0$, $\lambda_0$, and $\dot{\lambda}_0$, thus the phase space of physical degrees of freedom is dimension four. Following the same procedure employed in the last section, the choice of canonical variables is
\begin{eqnarray}
Q_1\equiv q   \longleftrightarrow  P_1&\equiv& \frac{\delta L}{\delta \dot{q}}=-(\gamma+\lambda) q^{(3)}-\dot{\lambda}\ddot{q}\nonumber \\
&&-[\lambda+\gamma(w_1^2+w_2^2)]\dot{q} \\
Q_2\equiv\dot{q}  \longleftrightarrow  P_2&\equiv& \frac{\delta L}{\delta \ddot{q}}=(\gamma+\lambda) \ddot{q}\label{20}\\
Q_3\equiv \lambda   \longleftrightarrow P_3&\equiv& \frac{\delta L}{\delta \dot{\lambda}}=0.
\end{eqnarray}

From eq. (\ref{20}), we can invert $\ddot{q}=P_2/(\gamma+Q_3)$, and the total Hamiltonian is
\begin{eqnarray}
H_{PUCT}&=&P_1Q_2+\frac{P_2^2}{2(\gamma+Q_3)}+\frac{1}{2}[Q_3+\gamma(w_1^2+w_2^2)]Q_2^2\nonumber \\
 && -\frac{\gamma}{2}w_1^2w_2^2Q_1^2+u_1\Phi_1.
\end{eqnarray} 
The primary constraint is $\Phi_1: P_3=0$, and there is only one secondary constraint
\begin{eqnarray}
\dot{\Phi}_1&=& [\Phi_1,H_T]_P = \frac{1}{2}\left[\frac{P_2^2}{(\gamma+Q_3)^2}-Q_2^2\right]\nonumber \\
\Rightarrow\Phi_2&:&\frac{P_2}{(\gamma+Q_3)}\pm Q_2\approx 0.
\end{eqnarray}
Here the constraint algorithm bifurcates, and we choose $P_2/(\gamma+Q_3)-Q_2\approx 0$ over $P_2/(\gamma+Q_3)+Q_2\approx 0$ --- one can check that choosing the other branch does not change the results.\footnote{Bifurcation simply means that there exist more than one constraint surface associated with the same variable. Operationally one chooses a bifurcation by specifying initial conditions.} The constraints are both second class and we can use them to rewrite $Q_3$, $P_3$ as functions of other canonical variables. The reduced Hamiltonian of the Pais-Uhlenbeck model with primary constraint $\ddot{q}^2-\dot{q}^2$ is thus
\begin{equation}
H_{PUCR}=P_1Q_2+P_2Q_2+\frac{\gamma}{2}(w_1^2+w_2^2-1)Q_2^2-\frac{\gamma}{2}w_1^2w_2^2Q_1^2.
\end{equation} 
This Hamiltonian remains linearly dependent on $P_1$ and $P_2$, hence it still suffers from the instability.

\section{$N$-th order theory with $M$ auxiliary variables} \label{sect:generalproof}

It is straightforward to generalize our result from the previous section to an $N$-th order derivative theory (with $N>2$) and $M$ auxiliary variables.

When we introduce constraints with $M$ auxiliary variables into an $N$-th order theory, it is clear that since the $M$ variables are non-dynamical, they will not enlarge the effective dimensionality of the original unconstrained phase space which is $2N$.  We consider the case where the number of constraints generated by $M$ auxiliary variables is $2M$ in section \ref{sect:2M}. 

\subsection{$M$ auxiliary variables with $2M$ constraints} \label{sect:2M}

Consider the most general $N$-th order theory with $M$ auxiliary variables
\begin{equation}
L_N=f(q, \dot{q},\ddot{q}, \dots, q^{(N)}, \lambda_1, \lambda_2, \dots, \lambda_M).
\end{equation}

There are $M+1$ Euler-Lagrange equations from varying $L_N$ with respect to $\lambda_a$ and $q$
\begin{eqnarray}
\frac{\partial f}{\partial \lambda_a}&=&0 (a=1,2,\dots, M)\\
\sum^N_{i=0}(-\frac{d}{dt})^i\frac{\partial f}{\partial q^{(i)}}&=&0 (i=0,1,2,\dots, N).
\end{eqnarray}

The total (unconstrained) phase space we start from is $2(N+M)$ dimensional, and the canonical variables are chosen as follows
\begin{eqnarray}
Q_1\equiv q  &\longleftrightarrow& P_1\equiv \sum^{N}_{j=1} (-\frac{d}{dt})^{j-1}\frac{\partial f}{\partial q^{(j)}} \label{eqn:canQ1} \\
&\vdots&\nonumber\\
Q_i\equiv q^{(i-1)} &\longleftrightarrow&  P_i\equiv \sum^{N}_{j=i} (-\frac{d}{dt})^{j-i}\frac{\partial f}{\partial q^{(j)}}\\
&\vdots&\nonumber \\
Q_N\equiv q^{(N-1)} &\longleftrightarrow&  P_N\equiv \frac{\partial f}{\partial q^{(N)}}\\
\nonumber\\
Q_{N+1}\equiv \lambda_1 &\longleftrightarrow& P_{N+1}\equiv P_{\lambda_1}=0\\
&\vdots& \nonumber \\
Q_{N+M}\equiv \lambda_M&\longleftrightarrow&  P_{N+M}\equiv P_{\lambda_M}=0. \label{eqn:canQNM}
\end{eqnarray}

Non-degeneracy means we can solve for $q^{(N)}$ as a function of $P_N$ and $Q_i$, i.e. $q^{(N)}=h(Q_1, \dots, Q_N, Q_{N+1}, \dots, Q_{N+M}, P_N)$. The total Hamiltonian takes the form
\begin{eqnarray}
H_T&=&P_1Q_2+\dots+P_{N-1}Q_N+P_N h(Q_1, \dots, Q_{N+M}, P_N)\nonumber \\
&&-f(Q_1, \dots, Q_{N+M}, h)+u_a\Phi_a,
\end{eqnarray}
where $\Phi_a: P_{N+a}=0$ are $M$ primary constraints, with $1\leq a \leq M$. We use the consistency relation to find the associated secondary constraints
\begin{equation}
\tilde{\Phi}_a = [\Phi_a,H_T]_P : \left.\frac{\partial f}{\partial \lambda_a}\right|_{\lambda_a=Q_{N+a}}\approx 0. \label{eqn:phiAcon}
\end{equation}

If $[\Phi_a, \tilde{\Phi}_b]_P\not\approx 0$ for $1\leq a, b\leq M$,  both $\Phi_a$ and $\tilde{\Phi}_b$ are second class constraints and thus there are no further constraints we can find using consistency relations --- we will consider the case when further constraints are present in the next section.  We can reduce M pairs of canonical variables $Q_{N+a}$, $P_{N+a}$ by using the constraints, i.e. $Q_{N+a}=F_a(Q_1,\dots, Q_N, P_N)$, $P_{N+a}=0$ and the reduced Hamiltonian on the $2N$ phase space becomes
\begin{eqnarray}
&&H_R=P_1Q_2+\dots+P_{N-1}Q_N\nonumber \\
&& +P_N h(Q_1,\dots, Q_N, F_a,\dots ,F_M, P_N)\nonumber \\
&& -f(Q_1, \dots, Q_N, h, F_a,\dots ,F_M)
\end{eqnarray}
which is linearly dependent on $P_1, \dots, P_{N-1}$ and thus is necessarily unstable. Therefore, we conclude that the Ostrogradski's instability survives if the auxiliary variables do not introduce enough constraints to reduce the dimensionality of the phase space. Since each auxiliary variable generates here only a pair of constraints, the dimensionality of the reduced phase space is the same as the one for the original theory without constraints
\begin{equation}
\mathrm{Total~}2(N+M)-2M~\mathrm{Constraints} = 2N.
\end{equation}

An example of this case is considered in section \ref{sect:PUqddot} above.

\section{Exorcising Ostrogradski's ghost by reducing the dimensionality of phase space} \label{sect:remove}
In the last section we showed that the Ostrogradski's ghost could not be exorcised if the effective dimensionality is not reduced, and in this section, we demonstrate such a reduction can render the theory stable. We will first introduce an example of higher derivative theory which is stabilized by the constraints, then we will show under what general conditions such a stabilization can occur.

\subsection{A simple example of stable non-degenerate higher derivative theory}\label{sect:example1}
Consider the following Lagrangian
\begin{equation}
L=\frac{\dot{q}^2}{2}+\frac{(\ddot{q}-\lambda)^2}{2},
\end{equation}
which is a non-degenerate higher derivative Lagrangian but is secretly stable as we will now show. As usual, the canonical variables are defined by 
\begin{eqnarray}
Q_1\equiv q   &\longleftrightarrow&   P_1\equiv \dot{q}-q^{(3)}+\dot{\lambda}\\
Q_2\equiv\dot{q}   &\longleftrightarrow&  P_2\equiv \ddot{q}-\lambda\\
Q_3\equiv \lambda   &\longleftrightarrow&  P_3\equiv 0,
\end{eqnarray}
where $\Phi_1:P_3=0$ is the primary constraint, and the total Hamiltonian is 
\begin{eqnarray}
H_T&=&P_1Q_2+P_2Q_3+\frac{P_2^2}{2}-\frac{Q_2^2}{2}+u_1\Phi_1.
\end{eqnarray}
The secondary constraints are again generated by the consistency relation $\dot{\Phi}_i\equiv[\Phi_i,H_T]\approx 0$, the secondary constraints of the theory are thus
\begin{eqnarray}
\Phi_2&:& \ \ \ \  \ \ \ -P_2\approx 0\\
\Phi_3&:&  \ \ P_1-Q_2\approx 0\\
\Phi_4&:&  -Q_3-P_2\approx 0.
\end{eqnarray}
One can check that all the constraints are second class constraints. Now, if we use ($\Phi_1$, $\Phi_4$) to reduce $(Q_3, P_3)$ and use ($\Phi_2$, $\Phi_3$) to reduce $(Q_2, P_2)$, the reduced Hamiltonian will become
\begin{eqnarray}
H_R&=&P_1^2-\frac{P_1^2}{2}=\frac{P_1^2}{2},
\end{eqnarray}
which is bounded from below and hence is free of the ghost. One can see the effective dimensionality of phase space is reduced from four ($Q_1,Q_2,P_1, P_2$) to two ($Q_1,P_1$). 

\subsection{General condition of stable non-degenerate higher derivative theory}
It turns out that the above procedure is not general --- a willy-nilly reduction of the phase space does not necessary lead to a stable theory. In this section we will find the condition such that the ghost is removed.

Considering the most general second order derivative theory with an auxiliary field $\lambda$, 
\begin{equation}
L=A_{ij}\lambda^i\ddot{q}^j,
\end{equation}
where $A_{ij}$ are functions of $q$ and $\dot{q}$. Note that here, we have used the subscripts on functions, i.e. $A_{ij}$ to \emph{label} the functions, while superscripts on variables, i.e. $\lambda$ denote its \emph{power}. We will use Einstein summation convention. For simplicity we restrict ourselves to the case where the auxiliary fields are at most quadratic, $i,j=0,1,2$, which guarantees a one to one mapping from the configuration space to the phase space. Again we follow the Dirac's analysis of constrained systems, by defining the canonical variables

\begin{eqnarray}
Q_1\equiv q   &\longleftrightarrow&   P_1\equiv \frac{\delta L}{\delta \dot{q}}=\frac{\partial L}{\partial \dot{q}}-\frac{d}{dt}\frac{\partial L}{\partial\ddot{q}}\\
Q_2\equiv\dot{q}   &\longleftrightarrow&  P_2\equiv A_{i1}\lambda^i+2A_{i2}\lambda^i\ddot{q}\label{conp2}\\
Q_3\equiv \lambda   &\longleftrightarrow&  P_3\equiv 0.
\end{eqnarray}
We can invert $\ddot{q}$ in r.h.s. of the eq.~(\ref{conp2}) as a function of canonical variables
\begin{eqnarray}
\ddot{q}\equiv h(Q_1,Q_2,Q_3, P_2)=\frac{P_2-A_{i1}Q_3^i}{2A_{j2}Q_3^j}.
\end{eqnarray}
The total Hamiltonian is thus
\begin{eqnarray}
H_T&=&P_1Q_2+P_2h(Q_1,Q_2,Q_3, P_2)+u_1\Phi_1\nonumber \\
&-&A_{ij}(Q_1,Q_2)Q_3^ih^j(Q_1,Q_2,Q_3, P_2)\label{contotal},
\end{eqnarray}
where $\Phi_1: P_{3}=0$ is the primary constraint of this theory and it generates a secondary constraint $\Phi_2$ by the consistency relation,
\begin{eqnarray}
&&\dot{P_3}\equiv[P_3, H_T]_{PB}\approx 0\nonumber\\
&\Rightarrow&- \left[P_2-\sum_{k,l} lA_{kl}Q_3^kh^{l-1}\right]\frac{\partial h}{\partial Q_3}
 +\sum_{i,j}iA_{ij}Q_3^{i-1}h^j\approx 0 
\nonumber \\
&\Rightarrow&\Phi_2: \sum_{i,j} iA_{ij}Q_3^{i-1}h^j=A_{1j}h^j+2A_{2j}\lambda h^j\approx 0.\label{conphi2}
\end{eqnarray}
From the second to the third weak equality the coefficient of $\partial h/\partial Q_3$ vanishes, by virtue of  eq.~(\ref{conp2}). 
 
To render the theory stable requires a reduction in the dimensionality of the original phase space. To ensure this, the consistency relations must continue to generate constraints beyond the first pair, which algebraically requires $\Phi_2$ to be independent of $Q_3$. The stable theory hence needs to obey the condition $\partial \Phi_2/\partial Q_3=0$, i.e.
\begin{eqnarray}
\frac{\partial \Phi_2}{\partial Q_3}&=&2A_{2i}h^i+jkA_{jk}Q_3^{j-1}h^{k-1}\frac{\partial h}{\partial Q_3}=0\label{conphi21}
\end{eqnarray}
where
\begin{align}
\frac{\partial h}{\partial Q_3}=-\frac{1}{2(A_{i2}\lambda^i)^2}[&(A_{j2}\lambda^j)(A_{11}+2A_{21}\lambda)\nonumber \\
&+(P_2-A_{k1}\lambda^k)(A_{12}+2A_{22}\lambda)].\label{conh}
\end{align}
 From eqs.~(\ref{conphi21})-(\ref{conh}), one can see $\partial \Phi_2/\partial Q_3$ is a quadratic function of $P_2$. To have vanishing $\partial \Phi_2/\partial Q_3$ we ask the coefficients of $P_2^0$, $P_2^1$, and $P_2^2$ to be zero.  This leads to the following most general conditions on $A_{ij}$ one can have with $\Phi_2$ independent of $Q_3$,
\[
 A_{ij} = \begin{bmatrix}
       A&B&a\\
c&\pm\sqrt{4ab}&0\\
b&0&0
     \end{bmatrix},
\]
where $A,B,a,b,c$ are all functions of $Q_1$ and $Q_2$. Furthermore $a$ is nonvanishing by construction or else the Lagrangian will not describe a higher derivative theory. The most general Lagrangian with more than two constraints now can be written as
\begin{equation}
L=A+B\ddot{q}+a\ddot{q}^2+c\lambda+b\lambda^2\pm\sqrt{4ab}\lambda \ddot{q}
\end{equation}
where all the coefficients are functions of $q$ and $\dot{q}$, and the ``acceleration" $\ddot{q}$ can be inverted by the definition of canonical momentum $P_2$ using eq.~(\ref{conp2})
\begin{eqnarray}
\ddot{q}=h=\frac{P_2-B\mp\sqrt{4ab}Q_3}{2a}.
\end{eqnarray}
The total Hamiltonian eq.(\ref{contotal}) and the secondary constraint $\Phi_2$ eq.(\ref{conphi2}) now can be rewritten as
\begin{eqnarray}
H_T&=&P_1Q_2+P_2h-A-Bh-ah^2-cQ_3\nonumber \\
&&-bQ_3^2\mp\sqrt{4ab}Q_3 h+u_1\Phi_1,\label{contotal2}\\
\Phi_2&:& c\pm \sqrt{4ab}(\frac{P_2-B}{2a})\approx 0\label{conphi22}.
\end{eqnarray}
Since the instability comes from the linear term $P_1Q_2$, to fix this instability we need to generate a constraint where $Q_2$ must be some function of $P_1$. To have a nontrivial theory, we need $P_1$ to enter the constraint equations either at $\Phi_3$ or $\Phi_4$.\footnote{If $P_1$ enters the constraint equations $\Phi_5$ or $\Phi_6$, there will be six constraints and by those constraints all the canonical variables will be some constants, and thus a trivial theory.} We will now show that the latter condition will not lead to a stable theory, and then show the condition for the former to lead to stability.

\subsubsection{$P_1$ entering in $\Phi_4$ does not lead to a stable theory}
To pick up $P_1$ in the constraint $\Phi_4$ requires $P_2$ to be in $\Phi_3$ but not before, i.e. $\Phi_2$ has to be independent of $P_2$. This can be achieved by specifying $b=0$ such that $\Phi_2:c\approx 0$ and $h=(P_2-B)/2a$.
Using the consistency relation, $\Phi_3$ is thus
\begin{equation}
\Phi_3: \frac{\partial c}{\partial Q_1}Q_2+\frac{\partial c}{\partial Q_2}\frac{(P_2-B)}{2a}.
\end{equation}
If $\partial c/\partial Q_2=0$, we will not be able to pick up $P_1$ at $\Phi_4$, which means the reduced Hamiltonian is either unstable (no constraint picks $P_1$ up) or trivial (theory with six constraints, all the variables are some constants). We thus require $\partial c/\partial Q_2\neq0$ in order to have a possibly stable theory with $P_1$ appearing in $\Phi_4$. One will see this requirement is also the same requirement for $Q_3$ to be in $\Phi_4$, since $\Phi_4$ can be generated again by consistency relation
 \begin{eqnarray}
\Phi_4&:&-\frac{\partial\Phi_3}{\partial P_2}(P_1-\frac{\partial A}{\partial Q_2}-\frac{\partial B}{\partial Q_2}h-\frac{\partial a}{\partial Q_2}h^2-\frac{\partial c}{\partial Q_2}Q_3) \nonumber \\
&&+\frac{\partial\Phi_3}{\partial Q_1}Q_2+\frac{\partial\Phi_3}{\partial Q_2}h \approx0 
\end{eqnarray}
Using $\Phi_1$, $\Phi_3$, $\Phi_4$ to eliminate $P_3$, $P_2$ and $Q_3$ and then substituting them into the total Hamiltonian~eq.({\ref{contotal2}}), the semi-reduced Hamiltonian becomes
\begin{equation}
H_{SR}=F(Q_1,Q_2)+P_1Q_2.
\end{equation}
If we substitute the last constraint $\Phi_2=c\approx 0$ which relates $Q_2$ to some function of $Q_1$, we will have the final reduced Hamiltonian
\begin{equation}
H_{R}=F_1(Q_1)+P_1F_2(Q_1),
\end{equation}
where $F_1$, $F_2$ are functions of $Q_1$ only. It is clear the final reduced Hamiltonian is always unstable unless $F_2=0$, that implies $c=Q_2F_3(Q_1)$ which means the Lagrange multiplier constrains $Q_1$ to be a constant and the theory is thus trivial. We conclude that if we want $P_1$ to appear only in the constraint $\Phi_4$, the theory is either unstable or trivial.

\subsubsection{$P_1$ entering at $\Phi_3$ and conditions for stability}
Finally we consider the case where $P_1$ enters at $\Phi_3$. This means that $P_2$ enters at $\Phi_2$ which requires that $b\neq 0$. Replacing $h$ in the total Hamiltonian eq.~(\ref{contotal2}), we get

\begin{equation}
H_T=P_1Q_2+\frac{(P_2-B\mp\sqrt{4ab}Q_3)^2}{4a}-A-cQ_3-bQ_3^2+u_1\Phi_1
\end{equation}
which we can use to calculate the awkward looking $\Phi_3$ 
\begin{eqnarray}
\Phi_3&:&\pm(P_2-B)\left((P_2-B)\frac{\partial b}{\partial Q_2}-2bQ_2\frac{\partial a}{\partial Q_1}\right)\nonumber \\
&&\pm a\left(2Q_2(P_2-B)\frac{\partial b}{\partial Q_1}-4b(P_1-\frac{\partial A}{\partial Q_2}+Q_2\frac{\partial B}{\partial Q_1})\right)\nonumber \\
&&+2(P_2-B)\sqrt{ab}\frac{\partial c}{\partial Q_2}+4aQ_2\sqrt{ab}\frac{\partial c}{\partial Q_1}\approx 0
\end{eqnarray}
which is always independent of $Q_3$, and because of $a, b\neq 0$, we can use $\Phi_3$ to express $P_1$ as other canonical variables on the constraint surface,
\begin{eqnarray}
P_1&\approx&\pm\left(\frac{c \ Q_2}{2\sqrt{ab}}\frac{\partial a}{\partial Q_1}-\frac{c\ Q_2\sqrt{ab}}{2b^2}\frac{\partial b}{\partial Q_1}+\frac{a\ Q_2}{\sqrt{ab}}\frac{\partial c}{\partial Q_1}\right)\nonumber\\
&&+\frac{\partial A}{\partial Q_2}-Q_2\frac{\partial B}{\partial Q_1}+\frac{c^2}{4b^2}\frac{\partial b}{\partial Q_2}-\frac{c}{2b}\frac{\partial c}{\partial Q_2}.\label{conp1}
\end{eqnarray}
If we use $\Phi_1$ and $\Phi_2$ to eliminate $P_3$ and $P_2$ in the total Hamiltonian, we can write semi-reduced Hamiltonian as
\begin{equation}
H_{SR}=P_1Q_2+\frac{c^2}{4b}-A.\label{consr}
\end{equation}
The last step for finding a stable final reduced Hamiltonian is to reverse eq.(\ref{conp1}) as $Q_2=g(Q_1, P_1)$ and substitute it into eq.(\ref{consr}). Since there are five arbitrary functions $A, B, a, b, c$,  we simply have to choose them as functions of $q$ and $\dot{q}$ such that the reduced Hamiltonian is stable. For example, in section~\ref{sect:example1} we chose $A=Q_2^2/2$, $B=c=0$, and $a=b=1/2$. 
\section{Summary} \label{sect:summary}

We prove that the linear instability, i.e. Ostrogradski's ghost, in a non-degenerate higher derivative theory can be exorcised by the addition of constraints, at the price of a reduction in the dimensionality of the phase space. We show this procedure in a class of second order time derivative theories with one Lagrange multiplier to illustrate how this is possible in principle. Generalization to arbitrary higher order derivative theory with multiple Lagrange multipliers is straightforward.

\appendix
\section{An example of stable constrained non-degenerate Pais-Uhlenbeck oscillator}\label{sect:example2}
The instability in the Pais-Uhlenbeck model can also be removed by introducing constraint in a way that the dimensionality of phase space is reduced. In this appendix, we consider the Lagrangian of Pais-Uhlenbeck model with the auxiliary field $\lambda$
\begin{eqnarray}
L&=&\frac{\gamma}{2}[\ddot{q}^2-(w_1^2+w_2^2)\dot{q}^2+w_1^2w_2^2q^2]\nonumber \\
&&+4\gamma w_1^2w_2^2q^2\lambda(1+\lambda)+2\sqrt{2}\gamma w_1w_2\lambda q\ddot{q},
\end{eqnarray}
the canonical variables are defined by
\begin{eqnarray}
Q_1\equiv q   \longleftrightarrow   P_1&\equiv& -\gamma[q^{(3)}+2\sqrt{2}w_1w_2(\lambda \dot{q}+\dot{\lambda}q)]\nonumber\\
&&-\gamma(w_1^2+w_2^2)\dot{q}\\
Q_2\equiv\dot{q}   \longleftrightarrow  P_2&\equiv& \gamma\ddot{q}+2\sqrt{2}\gamma w_1w_2\lambda q\\
Q_3\equiv \lambda   \longleftrightarrow  P_3&\equiv& 0,
\end{eqnarray}
where $\Phi_1:P_3=0$ is the primary constraint, and the total Hamiltonian is 
\begin{eqnarray}
H_T&=&P_1Q_2+\frac{P_2^2}{2\gamma}-\frac{\gamma}{2}w_1^2w_2^2Q_1^2+\frac{\gamma}{2}(w_1^2+w_2^2)Q_2^2\nonumber\\
&&-4\gamma w_1^2w_2^2Q_1^2Q_3-2\sqrt{2}w_1w_2Q_1Q_3P_2+u_1\Phi_1.\nonumber \\
\end{eqnarray}
The secondary constraints generated by the consistency relation are thus
\begin{eqnarray}
\Phi_2&:&  \ \ \ \ \ \ \ \ \ \ \ \ \ \ \ \ \ \sqrt{2}\gamma w_1w_2Q_1+P_2\approx 0\\
\Phi_3&:&  P_1+\gamma(w_1^2+w_2^2-\sqrt{2}w_1w_2)Q_2\approx 0\\
\Phi_4&:&  w_1w_2(3+8Q_3)-\sqrt{2}(w_1^2+w_2^2)(1+2Q_3)\approx 0.\nonumber\\
\end{eqnarray}
One can check all the constraints are second class constraints. Now if we use ($\Phi_1$, $\Phi_4$) to reduce $(Q_3, P_3)$ and use ($\Phi_2$, $\Phi_3$) to reduce $(Q_2, P_2)$, the reduced Hamiltonian will become
\begin{eqnarray}
H_R&=&\frac{\gamma}{2}w_1^2w_2^2Q_1^2+\frac{w_1w_2P_1^2}{\sqrt{2}\gamma(\sqrt{2}w_1w_2-w_1^2-w_2^2)^2},\nonumber \\
\end{eqnarray}
which is positive definite. One can see the effective dimensionality of phase space is reduced from four ($Q_1,Q_2,P_1, P_2$) to two ($Q_1,P_1$).

\acknowledgments

We would like to thank Kurt Hinterbichler, Alex Vikman, Ignacy Sawicki, Richard Woodard, Chris Pope, Richard Easther, Peter Adshead, Thomas Sotiriou, Pei-Ming Ho and Neil Barnaby for useful conversations. TJC acknowledges funding from the Cambridge Overseas Trust and A.J.T. was supported in part by the Department of Energy under grant DE-FG02-12ER41810.

\end{document}